\documentclass[preprint,12pt]{elsarticle}

\usepackage{array}
\usepackage{threeparttable}

\usepackage{amsmath}
\usepackage{multirow}
\usepackage{booktabs}
\usepackage{graphicx}
\usepackage{tabularx}
\usepackage{float}
\usepackage{multicol}
\usepackage{multirow}
\usepackage{stfloats} 

\newcommand*{\Scale}[2][4]{\scalebox{#1}{$#2$}}%

\newcounter{appcount}
\newcommand{\appendicesname}
            {Appendix\ \thechapter  \Alph{appcount}}
\newcommand{\bookappendicesname}
            {Appendix\ \Alph{appcount}}
\newcommand{\chapterappendix}[1]
          {\par\setcounter{section}{0}
           \setcounter{equation}{0}
           \setcounter{table}{0}
           \setcounter{figure}{0}
          \addtocounter{appcount}{1}   \renewcommand{\theequation}{\thechapter\Alph{appcount}.\arabic{equation}}
          \renewcommand{\thetable}{\thechapter\Alph{appcount}.\arabic{table}}
          \renewcommand{\thefigure}{\thechapter\Alph{appcount}.\arabic{figure}}
           \setcounter{section}{\arabic{chapter}\Alph{section}}
           \if@openright\cleardoublepage\else\clearpage\fi
           \chapter*{\huge{\appendicesname}\newline\newline \Huge{#1}}
           \addcontentsline{toc}{section}{\thechapter\Alph{appcount} #1}
           \markright{\MakeUppercase{\appendicesname.\ { #1}}}}
\newcommand{\bookappendix}[1]
          {\par\setcounter{section}{0}
           \setcounter{equation}{0}
           \setcounter{table}{0}
           \setcounter{figure}{0}
          \addtocounter{appcount}{1}   \renewcommand{\theequation}{\Alph{appcount}.\arabic{equation}}
           \renewcommand{\thetable}{\Alph{appcount}.\arabic{table}}
             \renewcommand{\thefigure}{\Alph{appcount}.\arabic{figure}}
           \setcounter{section}{\arabic{chapter}\Alph{section}}
           \if@openright\cleardoublepage\else\clearpage\fi
           \chapter*{\huge{\bookappendicesname}\newline\newline \Huge{#1}}
           \addcontentsline{toc}{chapter}{\bookappendicesname #1}
          \markright{\MakeUppercase{\bookappendicesname.\ { #1}}} }
\newcounter{example}
\newcounter{property}
\newcommand{\ben}{\begin{equation}}
\newcommand{\een}{\end{equation}}
\newcommand{\bea}{\begin{eqnarray*}}
\newcommand{\eea}{\end{eqnarray*}}
\newcommand{\bean}{\begin{eqnarray}}
\newcommand{\eean}{\end{eqnarray}}

\newcommand{\bfalph}{\mbox{\boldmath{$\alpha$}}}

\newcommand{\bfth}{\mbox{\boldmath{$\theta$}}}

\journal{Signal Processing}

\begin{document}

\begin{frontmatter}

\title{Adaptive Signal Detection and Parameter Estimation in Unknown Colored Gaussian Noise}

\author{Bo Tang, Haibo He and Steven Kay}

\address{Bo Tang, Haibo He and Steven Kay are with the Department of Electrical, Computer, and Biomedical Engineering, University of Rhode Island, Kingston, RI, 02881 USA, Email: \textit{btang, he, kay}@ele.uri.edu. \\
Corresponding author: Haibo He (Email: he@ele.uri.edu).}

\begin{abstract}
This paper considers the general signal detection and parameter estimation problem in the presence of colored Gaussian noise disturbance. By modeling the disturbance with an autoregressive process, we present three signal detectors with different unknown parameters under the general framework of binary hypothesis testing. The closed form of parameter estimates and the asymptotic distributions of these three tests are also given. Given two examples of frequency modulated signal detection problem and time series moving object detection problem, the simulation results demonstrate the effectiveness of three presented detectors.
\end{abstract}


\end{frontmatter}


\section{Introduction}
Signal detection is widely used in many applications, including image and video processing, wireless communication, signal processing, and classification. Signal detection aims to detect whether a data observation contains the signal that is usually embedded with noise. For instance, a common radar or sonar problem is to detect target signal with unknown amplitude in the presence of noise. The successful detection results are critical for the next decision making of operators. 

Over the last few decades, many detectors have been designed in both scientific and engineering fields. Most of these detectors consider that the target signal is surrounded or disturbed by the white Gaussian noise. The assumption Gaussian noise offers many advantages for signal detection or recognition. In an adaptive detection procedure \cite{reed1974rapid}, the statistical properties of the Gaussian noise can be estimated by the signal that contains only noise from other returns. While only the covariance matrix of noise is unknown, Kelly derived a generalized likelihood ratio test (GLRT) rule for detection of signal of unknown amplitude \cite{kelly1986adaptive} by formulating it as hypothesis testing problem. The test exhibits the property of constant false alarm rate (CFAR) detector indicating that the false alarm rate is irrelevant to the unknown covariance matrix of the noise \cite{fuhrmann1992cfar}. The probability density function (PDF) of Kelly's GLRT detector was also derived. Later work further demonstrates that the GLRT detector is the uniformly most powerful invariant detector \cite{scharf1994matched}. For this problem of detecting a signal disturbed by Gaussian noise with unknown covariance matrix, Rao test detector also exhibits the CFAR behavior and has a matched detection performance as that of the GLRT detector if sufficient training data is available \cite{de2007rao}. 

However, the assumption of Gaussian noise may be not always true for many practical problems of interest. The colored Gaussian noise may degrade the detection performance of the existing GLRT and Rao test detectors devised for Gaussian noise \cite{kay1983asymptotically,tang2016detection,tangEEF}. To address this issue, many colored Gaussian noise models have been studied. 
Kay has derived a GLRT detector for detecting a known signal in \cite{kay1983asymptotically} and has studied the problem of parameters estimation for time series modeling using a parameter transformation in \cite{kay1994maximum}. 
More recently, the authors in \cite{mahmoudi2010parameter} and \cite{mahmoudi2012parameter} have attempted to address the issues of parameter estimation of autoregressive signals corrupted with colored noise. The existing works derive signal detectors or parameter estimators under certain conditions with particular unknown parameters. This is partly because some detectors may not exist for some cases (e.g., the GLRT detector when the autoregressive coefficients are unknown), and the researchers have to seek for the alternative one. 

In this paper, we extend Kay's work in \cite{kay1983asymptotically} and provide solutions for a general signal detection problem in the presence of colored Gaussian noise which is modeled via the autoregressive (AR) process. Three test detectors based upon GLRT and Rao test criteria are built by modeling the colored noise with an AR process when different parameters are unknown. We present the closed-form expressions of unknown parameter estimates using the maximum likelihood methods and the asymptotic distributions of these test detectors. Two examples of stepped-frequency signal detection and moving object detection are studied through computational simulations to illustrate wide practical applications of three given test detectors. 

The remaining paper is organized as follows: In section 2, we formulate the signal detection model in term of hypothesis testing. In section 3, we generalize the hypothesis testing problem for deterministic signal with AR model and build three sub-optimal test detectors. The exact performance of two GLRTs and the asymptotic performance of the Rao test are given. In section 4, we demonstrate the applicability of three detectors on stepped-frequency signal detection, and extend three detectors for time-series signal detection, followed by the conclusion in section 5. 

\section{Signal Model}
In radar, as well as in other applications, the classic signal detection problem is to detect whether a signal $\mathbf{s} = [s[0], s[1], \cdots, s[N-1]]^T$ with unknown amplitude $A$ appears (hypothesis $\mathcal{H}_1$) in an observation $\mathbf{x} = [x[0], x[1], \cdots, x[N-1]]^T$ or not (hypothesis $\mathcal{H}_0$), which is formulated as follows
\begin{align}
\label{hypothesis_object_detection}
& \mathcal{H}_0 : x[n] = w[n] & \nonumber \\
& \mathcal{H}_1 : x[n] = As[n]+ w[n] & n = 0,1,\cdots,N-1
\end{align}
where the observation $\mathbf{x}$ is disturbed by a colored noise vector $\mathbf{w}$. We model the colored noise with the following AR process
\begin{align}
\label{hypothesis_AR}
w[n] = \sum_{i=1}^p \alpha_i w[n-i] + v[n] \quad n = 0,1,\cdots,N-1
\end{align}
where $v[n]$ is assumed to be independent and identically distributed (I.I.D.) random variable and satisfies a Gaussian distribution, that is, we have $\mathbf{v} \sim \mathcal{N}(\mathbf{0}, \sigma^2 \mathbf{I})$ where $\mathbf{v} = [v[0], v[1], \cdots, v[N-1]]^T$ and $\mathbf{I}$ is a $N \times N$ identity matrix. 

For the binary hypothesis testing problem, the Neyman-Pearson (NP) rule which maximizes the detection accuracy subject to the constraint of false alarm provides the most optimal results when we assume the distributions of two hypotheses are completely known. But it is unpractical to know their distributions prior to the beginning of detection because of some unknown parameters, and thus the most optimal NP detector does not exist. The alternative choice is to build the sub-optimal detectors. For the binary hypothesis testing in Eq. (\ref{hypothesis_object_detection}) and (\ref{hypothesis_AR}), the following parameters maybe unknown, including the coefficient factors of the AR process $\alpha_1, \alpha_2, \cdots, \alpha_p$, the signal $\mathbf{s}$ or its amplitude $A$, and the variance of the white Gaussian noise $\sigma^2$. In this paper, we consider the following three cases and respectively build their sub-optimal detectors for signal detection: 1. when $\mathbf{s}$ is unknown; 2. when $\mathbf{s}$ and $\sigma^2$ are unknown; 3. when $\mathbf{s}$, $\sigma^2$ and $\alpha_1, \alpha_2, \cdots, \alpha_p$ are all unknown. 

\section{Adaptive Signal Detection for Colored Noise}

\subsection{Hypothesis Testing for Linear Signal Model}
Instead of building specific detector for different detection problem, most detection problems can be effectively solved by employing a general signal model. The solutions for classic or Bayesian linear model in which the parameters is assumed to be deterministic, are generalized in \cite{Kay:1998_detection} for a wide class of detection problems when the signal is corrupted by Gaussian noise. Following the work in \cite{Kay:1998_detection}, we now generalize the deterministic signal model in colored noise with an AR model. 

Firstly, we assume that the data has the form of $\mathbf{x} = \mathbf{H} \bfth + \mathbf{w}$, where $\mathbf{H}$ is a known $N \times q$ $(N > q)$ observation matrix, $\bfth$ is a $q \times 1$ parameter vector, and $\mathbf{w}$ is an $N \times 1$ noise vector modeled by an AR($p$) model in Eq. (\ref{hypothesis_AR}). 

Moreover, for the generality, we wish to test whether the parameters $\bfth$ satisfy the linear equation $\mathbf{A} \bfth = \mathbf{b}$ as opposed to $\mathbf{A} \bfth \neq \mathbf{b}$, where $\mathbf{A}$ is a $r \times q$ matrix ($r \leq q$) of rank $r$, $\mathbf{b}$ is a $r \times 1$ vector. The assumption of matrix $\mathbf{A}$ with rank $r$ ensures that there is only one solution for $\bfth$. Hence, the hypothesis testing problem for deterministic signal with the AR model is defined as
\begin{align}
\label{classic_hypo}
\left\{ 
\begin{array}{l}
\mathcal{H}_0: \quad \mathbf{A} \bfth = \mathbf{b} \\
\mathcal{H}_1: \quad \mathbf{A} \bfth \neq \mathbf{b} 
\end{array}
\right.
\end{align}

In this classic AR model with unknown deterministic signal parameters, a uniformly most powerful (UMP) test which aims to produce the highest probability of detection $P_D$ for all values of the unknown signal parameters given the probability of false alarm $P_{FA}$ does not always exist \cite{Kay:1998_detection}. Thus, the sub-optimal detectors with good detection performance are considered, such as the GLRT detector and the Rao test detector. Usually the detection loss in both GLRT and Rao test detector is quite small and their performances are bounded by the UMP detector if the perfect knowledge of unknown parameters are completely known.

\subsection{GLRT Detector with Unknown $\bfth$}
In many detection problems, the parameters of distribution under null hypothesis $\mathcal{H}_0$ (only noise), such as $\bfalph$ and $\sigma^2$, are statistically known. However, the signal $\bfth$ is not constant and unknown, such as object detection in radar and sonar and carrier signal detection in communication. For this case, we build the GLRT detector for unknown $\bfth$ in Theorem 1.

\newtheorem{theorem}{\bf Theorem}
\begin{theorem}
\label{theorem1}
Assume the data satisfies the classic AR model in Eq. (\ref{classic_hypo}), and the binary hypothesis testing problem with unknown $\bfth$ is defined in Eq. (\ref{classic_hypo}). We decide $\mathcal{H}_1$ if 
\begin{align}
\label{T1}
& T_{G_1}(\mathbf{x})  = 2\ln L_{G}(\mathbf{x}) = 2 \ln \frac{p(\mathbf{x}; \mathcal{H}_1)}{p(\mathbf{x}; \mathcal{H}_0)} \\
& = \frac{( \mathbf{A} \hat{\bfth}_1 - \mathbf{b})^{T} [ \mathbf{A} [ (\mathbf{TH})^{T} \mathbf{TH} ]^{-1} \mathbf{A}^{T} ]^{-1} ( \mathbf{A}\hat{\bfth}_1 - \mathbf{b})}{\sigma^2}  > \gamma'
\end{align}
where $\hat{\bfth}_1 = [(\mathbf{TH})^{T} \mathbf{TH}]^{-1}(\mathbf{TH})^{T}(\mathbf{Tx}+\mathbf{c})$, $\mathbf{T}$ is a $N \times N$ matrix and $\mathbf{c}$ is a $N \times 1$ vector. Both $\mathbf{T}$ and $\mathbf{c}$ are defined in Eq. (\ref{T_c_form}) in which $\alpha_0 = 1$, $\alpha_1, \alpha_2, \cdots, \alpha_p$ are known AR coefficients, $x[k]$ denotes the $k$-th element of vector $\mathbf{x}$, and $x[-1], x[-2], \cdots, x[-p]$ are given initial values.
\begin{eqnarray}
\label{T_c_form}
\Scale[0.75]{
\mathbf{T} = \left( \begin{array}{c c c c c c c c c}
&\alpha_0 & 0 & 0 & \cdots & \cdots & \cdots & 0 \\
&-\alpha_1 & \alpha_0 & 0 & \cdots & \cdots & \cdots& 0 \\
 &  & \ddots &   &  & & & \vdots \\
&-\alpha_p & -\alpha_{p-1} & \cdots & \alpha_0 & \cdots & \cdots & 0 \\
&0 & -\alpha_{p} & -\alpha_{p-1} & \cdots & \alpha_0 &  \cdots & 0 \\
 &  & \ddots &   &  & & & \vdots \\
&0 & \cdots & \cdots & -\alpha_{p} & -\alpha_{p-1} & \cdots & \alpha_0 \\
\end{array} 
\right)
\quad \text{and }
\mathbf{c} = \left( \begin{array}{c c }
&-\sum_{k=1}^{p} \alpha_k x[-k] \\
&-\sum_{k=2}^{p}\alpha_k x[-k] \\
&\vdots \\
&-\alpha_p x[-p] \\
&0\\
&\vdots \\
&0
\end{array} 
\right)
}
\end{eqnarray}

The $\gamma'$ is the threshold which can be determined by the probability of false alarm. The exact detection performance is given as follows
\begin{align}
\label{Theory 1 Performance}
& P_{FA} = Q_{\chi_{r}^2}(\gamma') \nonumber \\
& P_D = Q_{\chi_{r}'^{2}(\lambda)}(\gamma')
\end{align}
where $\chi_{r}^2$ denotes the central chi-squared distribution, $\chi_{r}'^{2}(\lambda)$ denotes the noncentral chi-squared distribution with the noncentrality parameter $\lambda$ which is
\begin{align}
\label{T1_lamda}
\lambda = \frac{( \mathbf{A} {\bfth}_1 - \mathbf{b})^{T} [ \mathbf{A} [ (\mathbf{TH})^{T} \mathbf{TH} ]^{-1} \mathbf{A}^{T} ]^{-1} ( \mathbf{A} {\bfth}_1 - \mathbf{b})}{\sigma^2}
\end{align}
where $\bfth_1$ is the true value of $\bfth$ under $\mathcal{H}_1$, and $Q(\gamma')$ function is its right-tail probability for a threshold $\gamma'$.
\end{theorem}

The proof of Theorem 1 is given in Appendix \ref{Appendix 1}. There are two useful remarks for this theorem: On one hand, for many detection problems, we try to test $\bfth = \mathbf{0}$ under $\mathcal{H}_0$ versus $\bfth \neq \mathbf{0}$ under $\mathcal{H}_1$, which is equivalent to test $\mathbf{s} = \mathbf{T} \mathbf{H} \bfth$ equals zeros or not. By setting $\mathbf{A} = \mathbf{I}$ and $\mathbf{b} = \mathbf{0}$ in the above detection theorem, this GLRT detector can be simplified as
\begin{align}
\label{theorem1_example}
T_{G_1}(\mathbf{x}) & = \frac{ \hat{\bfth}_1^{T} (\mathbf{TH})^{T} \mathbf{TH} \hat{\bfth}_1}{\sigma^2} \nonumber \\
& =  \frac{ \hat{\mathbf{s}}^{T} \hat{\mathbf{s}}}{\sigma^2} > \gamma'
\end{align}
where $\hat{\mathbf{s}} = \mathbf{TH}\hat{\bfth}_1$. Thus, the GLRT can be interpreted as energy detector, and further as estimator-correlator (see in \cite{Kay:1998_detection}). On the other hand, it is worth noticing that the detection performance as given by Eq. (\ref{Theory 1 Performance}) is exact for finite data records, which indicates that this GLRT detector could be applied for many problems and allows us to set the threshold as well as to determine the probability of detection.

\subsection{GLRT Detector with Unknown $\bfth$ and $\sigma^2$}
We furthermore consider the case that both $\bfth$ and $\sigma^2$ are unknown. The GLRT detector for deterministic signal with AR($p$) model is given in Theorem 2.

\begin{theorem}
\label{theorem2}
Assume the data from classic AR model in Eq. (\ref{classic_hypo}), and the binary hypothesis testing problem with unknown $\bfth$ and unknown $\sigma^2$ is defined in Eq. (\ref{classic_hypo}). We decide $\mathcal{H}_1$ if
\begin{align}
&T_{G_2}(\mathbf{x})  = \frac{N-q}{r} (L_G(\mathbf{x})^{\frac{2}{N}}-1) = \frac{N-q}{r} \times \nonumber \\
& \frac{( \mathbf{A} \hat{\bfth}_1 - \mathbf{b})^{T} [ \mathbf{A} [ (\mathbf{TH})^{T} \mathbf{TH} ]^{-1} \mathbf{A}^{T} ]^{-1} ( \mathbf{A}\hat{\bfth}_1 - \mathbf{b})}{( \mathbf{Tx} + \mathbf{c})^{T}[\mathbf{I}- \mathbf{TH} [(\mathbf{TH})^{T} \mathbf{TH}]^{-1} (\mathbf{TH})^{T}]( \mathbf{Tx} + \mathbf{c})}  > \gamma'
\end{align}
where $\hat{\bfth}_1 = [(\mathbf{TH})^{T} \mathbf{TH}]^{-1}(\mathbf{TH})^{T}(\mathbf{Tx}+\mathbf{c})$, $\mathbf{T}$ and $\mathbf{c}$ are given in Eq. (\ref{T_c_form}). The exact detection performance is given by
\begin{align}
& P_{FA} = Q_{F_{r,N-q}}(\gamma') \nonumber \\
& P_D = Q_{F_{r,N-q}'(\lambda)}(\gamma')
\end{align}
where $F_{r,N-q}$ denotes the central $F$ distribution, and $F_{r,N-q}'(\lambda)$ denotes the noncentral $F$ distribution with the noncentrality parameter $\lambda$ which is
\begin{align}
\lambda = \frac{( \mathbf{A} {\bfth}_1 - \mathbf{b})^{T} [ \mathbf{A} [ (\mathbf{TH})^{T} \mathbf{TH} ]^{-1} \mathbf{A}^{T} ]^{-1} ( \mathbf{A} {\bfth}_1 - \mathbf{b})}{\sigma^2}
\end{align}
where $\bfth_1$ is the true value of $\bfth$ under $\mathcal{H}_1$.
\end{theorem}

This theorem can be easily proved based on the Theorem 1 and the detection theory of classic linear model in \cite{Kay:1998_detection}. Similar to Theorem 1, this GLRT detector can be interpreted as the energy detector and estimator-correlator but both $\bfth$ and $\sigma^2$ need to be estimated under $\mathcal{H}_1$ and $\mathcal{H}_0$. Also the detection performance for finite data records is exactly given. Note that this GLRT detector is identical with the GLRT detector in Theorem 1, when the parameter $\sigma^2$ is known, otherwise, the $\sigma^2$ is replaced with its MLE estimate $\hat{\sigma}^2_1$ under $\mathcal{H}_1$
\begin{align}
\label{T2_estimatSig}
\hat{\sigma}^2_1 = \frac{\mathbf{x}^T(\mathbf{I} - \mathbf{H}(\mathbf{H}^T\mathbf{H})^{-1}\mathbf{H}^T)^{-1}\mathbf{x}}{N-q}
\end{align}

\subsection{Rao Test Detector with Unknown $\bfth$, $\sigma^2$ and $\alpha_1, \alpha_2, \cdots, \alpha_p$}
When $\bfth$, $\bfalph$ and $\sigma^2$ are all unknown, it is difficult to work with the exact expression of GLRT detector. Instead of GLRT detector, we build the Rao test detector for this case. This is because the Rao test only need to estimate the nuisance parameters under $\mathcal{H}_0$. For the test with nuisance parameters and with the assumptions that the signal is weak and the data record is large, Rao test is the asymptotically equivalent with GLRT test. Also, the asymptotic performance of Rao test is the same as GLRT.

\begin{theorem}
\label{theorem2}
Assume the data from classic AR model in Eq. (\ref{classic_hypo}) and define all unknown parameter as a vector $\bfth = [\bfth_r^T, \bfth_w^T]^T$. To avoid confusion in the definition of $\bfth$, we replace the unknown signal $\bfth$ used in the previous sections by the vector $\bfth_r$, and define the nuisance parameters $\bfth_w = [\bfalph \quad \sigma^2]$. The Rao test for the binary hypothesis testing problem becomes
\begin{align}
\label{case3}
& \mathcal{H}_0: \quad \mathbf{A} \bfth_r = \mathbf{b}, \bfth_w \nonumber \\
& \mathcal{H}_1: \quad \mathbf{A} \bfth_r \neq \mathbf{b}, \bfth_w 
\end{align} 
where the unknown nuisance parameters vector $\bfth_w$ is the same under either hypothesis $\mathcal{H}_0$ or $\mathcal{H}_1$. We decide $\mathcal{H}_1$ if 
\begin{align}
\label{T3}
& T_{R}(\mathbf{x}) = \left. \frac{\partial \ln p(\mathbf{x}; \bfth)}{\partial \bfth_{r}} \right |^{T}_{\bfth = \hat{\bfth}_0} \left[ \mathbf{I}^{-1}(\bfth)\right]_{rr} \left. \frac{\partial \ln p(\mathbf{x}; \bfth)}{\partial \bfth_{r}} \right |_{\bfth = \hat{\bfth}_0} \nonumber \\
& = \frac{1}{\hat{\sigma}^2} [(\hat{\mathbf{T}} \mathbf{x} + \mathbf{\hat{c}})^T \hat{\mathbf{T}} \mathbf{H} [(\hat{\mathbf{T}} \mathbf{H})^T \hat{\mathbf{T}} \mathbf{H}]^{-1} (\hat{\mathbf{T}} \mathbf{H})^T (\hat{\mathbf{T}} \mathbf{x} + \mathbf{\hat{c}}) \nonumber \\
& \quad - (\hat{\mathbf{T}} \mathbf{x} + \mathbf{\hat{c}})^T \hat{\mathbf{T}} \mathbf{H} \bfth_{r_0} - \bfth_{r_0}^T (\hat{\mathbf{T}} \mathbf{H})^T (\hat{\mathbf{T}} \mathbf{x} + \mathbf{\hat{c}}) \nonumber \\
& \quad + \bfth_{r_0}^T (\hat{\mathbf{T}} \mathbf{H})^T \hat{\mathbf{T}} \mathbf{H} \mathbf{A}^{-1} \mathbf{b}] > \gamma'
\end{align}
where $\hat{\bfth}_0 = [\bfth_{r_0}^T \hat{\bfth}^T_{w_0}]^T$, $\bfth_{r_0}$ is the solution of $\bfth_r$ under $\mathcal{H}_0$ and $\hat{\bfth}_{w_0}$ is the MLE of $\bfth_w$ under $\mathcal{H}_0$. The MLEs of $\sigma^2$ is given by
\begin{align}
\label{eq:Theory3_sigma_hat}
\hat{\sigma}^2 = \frac{(\mathbf{\hat{T}x + \hat{c}})^{T}(\mathbf{\hat{T}x + \hat{c}})}{N}
\end{align} 
and $\hat{\mathbf{T}}$ and $\hat{\mathbf{c}}$ are shown in Eq. (\ref{Theory3: T_c_hat_form}).

\begin{eqnarray}
\label{Theory3: T_c_hat_form}
\Scale[0.75]{
\hat{\mathbf{T}} = \left( \begin{array}{c c c c c c c c c}
&\hat{\alpha}_0 & 0 & 0 & \cdots & \cdots & \cdots & 0 \\
&-\hat{\alpha}_1 & \hat{\alpha}_0 & 0 & \cdots & \cdots & \cdots& 0 \\
 &  & \ddots &   &  & & & \vdots \\
&-\hat{\alpha}_p & -\hat{\alpha}_{p-1} & \cdots & \hat{\alpha}_0 & \cdots & \cdots & 0 \\
&0 & -\hat{\alpha}_{p} & -\hat{\alpha}_{p-1} & \cdots & \hat{\alpha}_0 &  \cdots & 0 \\
 &  & \ddots &   &  & & & \vdots \\
&0 & \cdots & \cdots & -\hat{\alpha}_{p} & -\hat{\alpha}_{p-1} & \cdots & \hat{\alpha}_0 \\
\end{array} 
\right)
\quad \text{and }
\mathbf{c} = \left( \begin{array}{c c }
&-\sum_{k=1}^{p} \hat{\alpha}_k x[-k] \\
&-\sum_{k=2}^{p}\hat{\alpha}_k x[-k] \\
&\cdots \\
&-\hat{\alpha}_p x[-p] \\
&0\\
&\cdots \\
&0
\end{array} 
\right)
}
\end{eqnarray}
where $\hat{\alpha_1},\hat{\alpha_2},\cdots,\hat{\alpha_p}$ are obtained by solving the following Yule-Walker equation \cite{Kay:1993_estimation}
\begin{eqnarray}
\label{eq:yw}
\Scale[0.75]{
\begin{array}{c c c c c c}
\begin{pmatrix}
r_{xx}[0] & r_{xx}[1] &  \cdots & r_{xx}[p-1] \\
r_{xx}[1] & r_{xx}[0] & \cdots & r_{xx}[p-2] \\
 & \ddots &  &   \vdots \\
r_{xx}[p-1] & r_{xx}[p-2] &  \cdots & r_{xx}[0] \\
\end{pmatrix}
\end{array} 
\begin{array}{c c}
\begin{pmatrix}
\hat{\alpha}_{1} \\
\hat{\alpha}_{2} \\
\vdots \\
\hat{\alpha}_{p} \\
\end{pmatrix}
\end{array} = \begin{array}{c c}
\begin{pmatrix}
\hat{r}_{xx}[0] \\
\hat{r}_{xx}[1] \\
\vdots \\
\hat{r}_{xx}[p-1] \\
\end{pmatrix}
\end{array} 
}
\end{eqnarray}


The asymptotic performance is given as follows
\begin{align}
\label{eq:Theory3_performance} 
& P_{FA} = Q_{\chi_{p}^2}(\gamma') \nonumber \\
& P_D = Q_{\chi_{p}'^2(\lambda)}(\gamma')
\end{align}
where the noncentrality parameter $\lambda$ is
\begin{align}
\lambda = \frac{\bfth_1^T (\mathbf{TH})^{T}\mathbf{TH} \bfth_1}{\sigma^2}
\end{align}
where $\bfth_1$ is the true value of $\bfth$ under $\mathcal{H}_1$.
\end{theorem}

The proof of Theorem 3 is given in Appendix \ref{Appendix 2}. Compared to the GLRT, the Rao test is easy to implement. More importantly, the Rao test would have a close detection performance when the signal is weak and the data record is large. For the finite data set, it would perform poorer than the GLRT detector. For the weak signal, we usually require large data records (large $N$) to improve the detection performance. While $N\rightarrow\infty$, the detection performance of the Rao test is the same as the GLRT. For the case $\mathbf{A} = \mathbf{I}$ and $\mathbf{b} = \mathbf{0}$, setting $\hat{\mathbf{x}}' = \mathbf{\hat{T}x} + \mathbf{\hat{c}}$, the Rao test in Eq. (\ref{T3}) can be simplified as
\begin{align}
T_{R}(\mathbf{x}) & = \frac{\hat{\mathbf{x}}'^T \mathbf{\hat{T} H} [(\mathbf{\hat{T} H})^{T} \mathbf{\hat{T} H}]^{-1}(\mathbf{\hat{T} H})^{T} \hat{\mathbf{x}}'}{\hat{\sigma}^2} > \gamma'
\end{align}


\subsection{Optimality Discussion}
\begin{figure}[!ht]
\centering
\includegraphics[width=8.6cm]{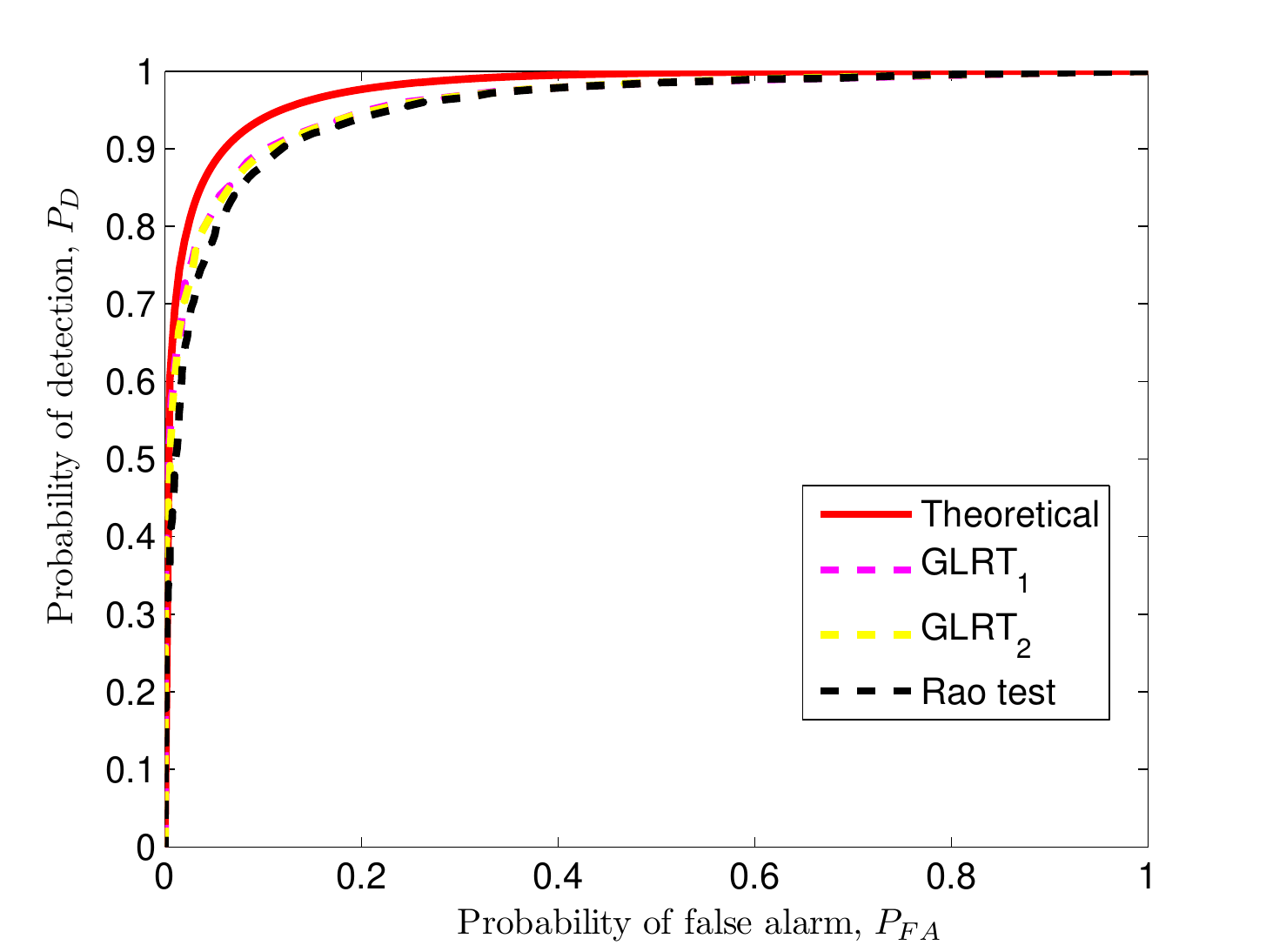}
\caption{Comparison of performance of three sub-optimal detectors to theoretical performance for DC level signal in AR noise.} 
\label{fig:ARsim}
\end{figure}

It is well known that the Bayesian detector (NP detector) is the most optimal one with the highest $P_D$ for a given $P_{FA}$ when the distributions under both $\mathcal{H}_0$ and $\mathcal{H}_1$ are completely known. If some of the parameters of distribution are unknown, the detection performance would be degraded because of the lack of information. For the detection problem with unknown parameters, the NP detector does not exist and we usually look for a sub-optimal detector, such as the GLRT detector or the Rao test detector, to obtain a good detection performance. It is worth noticing that the detection performance is decreasing with the loss of information. In other words, we cannot expect that a detector achieves better performance while more parameters are unknown. Hence, the detection performance of the above three presented detectors is degraded in comparison with the NP detector which provides the upper bound for them when all the model parameters are assumed to be known. To demonstrate this, we carry out one comparative experiment, where the signal to be detected has a DC level: $x[n] = A + w[n], n = 0, 1, \cdots, N-1$. We choose $A = 0.5$, $\alpha_1 = 0.5, \alpha_2 = -0.3$, $\sigma^2 = 2$, and $N = 100$. Three test detectors have been built under the above three respective cases. We show the simulation results in Fig. \ref{fig:ARsim}. The theoretical detector is the NP detector in which all the parameters are assumed to be completely known, and the detection performance with respect to the probability of false alarm is given by
\begin{align}
P_D = Q\left(Q^{-1}(P_{FA}) - \frac{A}{\sigma} \sqrt{(\mathbf{T}\mathbf{H})^{T} \mathbf{T}\mathbf{H}} \right)
\end{align}

The results in Fig. \ref{fig:ARsim} also illustrate that the GLRT detectors perform better than the Rao test detector because of more unknown parameters in case 3. Although the GLRT detector always performs very well for various detection problems, its analytic form is difficult to be determined for some cases, such as case 3 in which the coefficients of the AR model are unknown. The Rao test is, therefore, one good alternative.


\section{Some Typical Applications}
\subsection{Application of Radar Signal Detection}
In active sonar or radar, the use of stepped-frequency signal can obtain high or super-high range resolution. A well-developed application of stepped-frequency signal is the synthetic aperture radar (SAR). To identify two targets that are close together, we need transmit a wide bandwidth signal which is a linear frequency modulated (FM) chirp, which is usually given by
\begin{align}
f(t) = A \exp [ j 2 \pi ( f_0 t + \frac{1}{2} k_0 t^2 ) ] \qquad 0 \leq t \leq T_p
\end{align}
where $f_0$ is the start frequency in Hz, $k_0$ is the sweep rate in Hz/sec, and $T_p$ is the pulse duration in sec. Hence, the bandwidth of transmitted signal is about $k_0 T$ Hz. We consider the signal reflected by targets is embedded with colored noise. If one only considers the real in-phase component (I) in the discrete time with $N$ samples, we aim to detect the signal with the following binary hypothesis testing
\begin{align}
\label{sar_signal_detection}
& \mathcal{H}_0 : x[n] = w[n] & \nonumber \\
& \mathcal{H}_1 : x[n] = A \cos[2 \pi (f_0 n + \frac{1}{2} k_0 n^2) + \phi ] + w[n]
\end{align}
where $n = 0,1,\cdots,N-1$. The amplitude $A$ and the phase $\phi$ are unknown, and the noise $\mathbf{w}$ is modeled by an AR process. By defining $\bfth = [\theta_1 \ \theta_2]^T$ where $\theta_1 = A \cos \phi$ and $\theta_2 = -A \sin \phi$, we have $\mathbf{x} = \mathbf{H} \bfth + \mathbf{w}$ \cite{Kay:1998_detection}.

\begin{figure*}[bp]
\centering
\includegraphics[width=14cm]{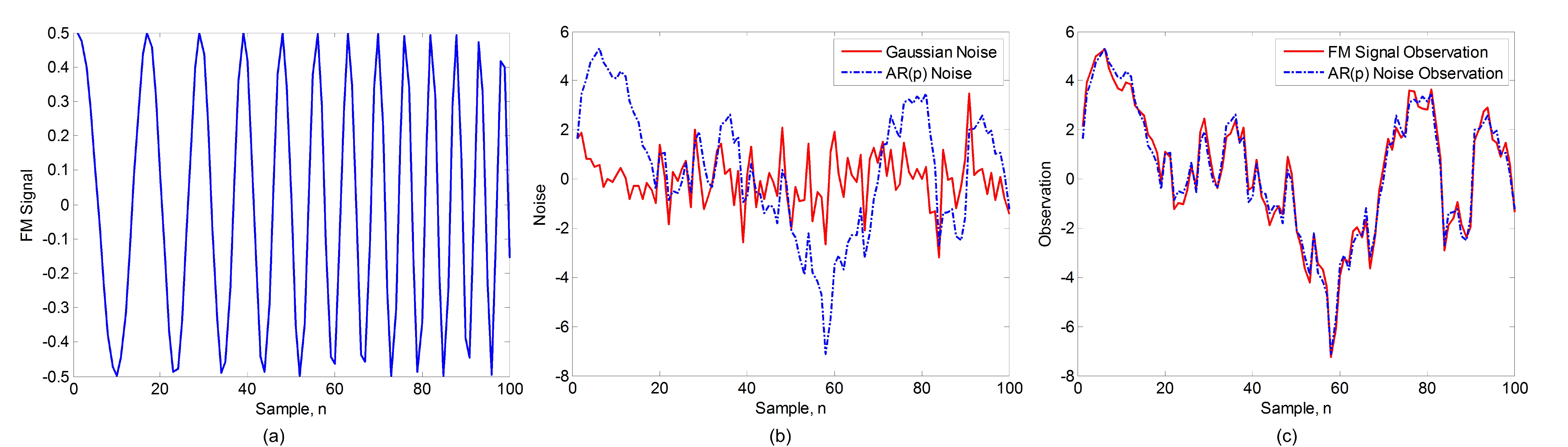}
\caption{FM signal and noise: (a) Linear FM signal (b) White Gaussian noise $v[n]$ with $\sigma^2 = 1$ and its colored AR($p$) noise $w[n]$ with $p=1$ and $\alpha_1 = -0.95$ (c) Observation of FM signal in the presence of colored noise and observation of only colored noise. } 
\label{fm_signal_noise}
\end{figure*}

We conducted experiments by applying three presented detectors to this signal detection problem when we consider different parameters are unknown. We followed the work in \cite{Kay:1998_detection} with the parameter setting: $A=0.5$, $f_0 = 0.05$, $k_0 = 0.0015$, $\phi = 0$, and $N = 100$. The AR process for noise modeling has one coefficient: $\alpha_1 = -0.95$. 

\begin{figure}[!ht]
\centering
\includegraphics[width=14cm]{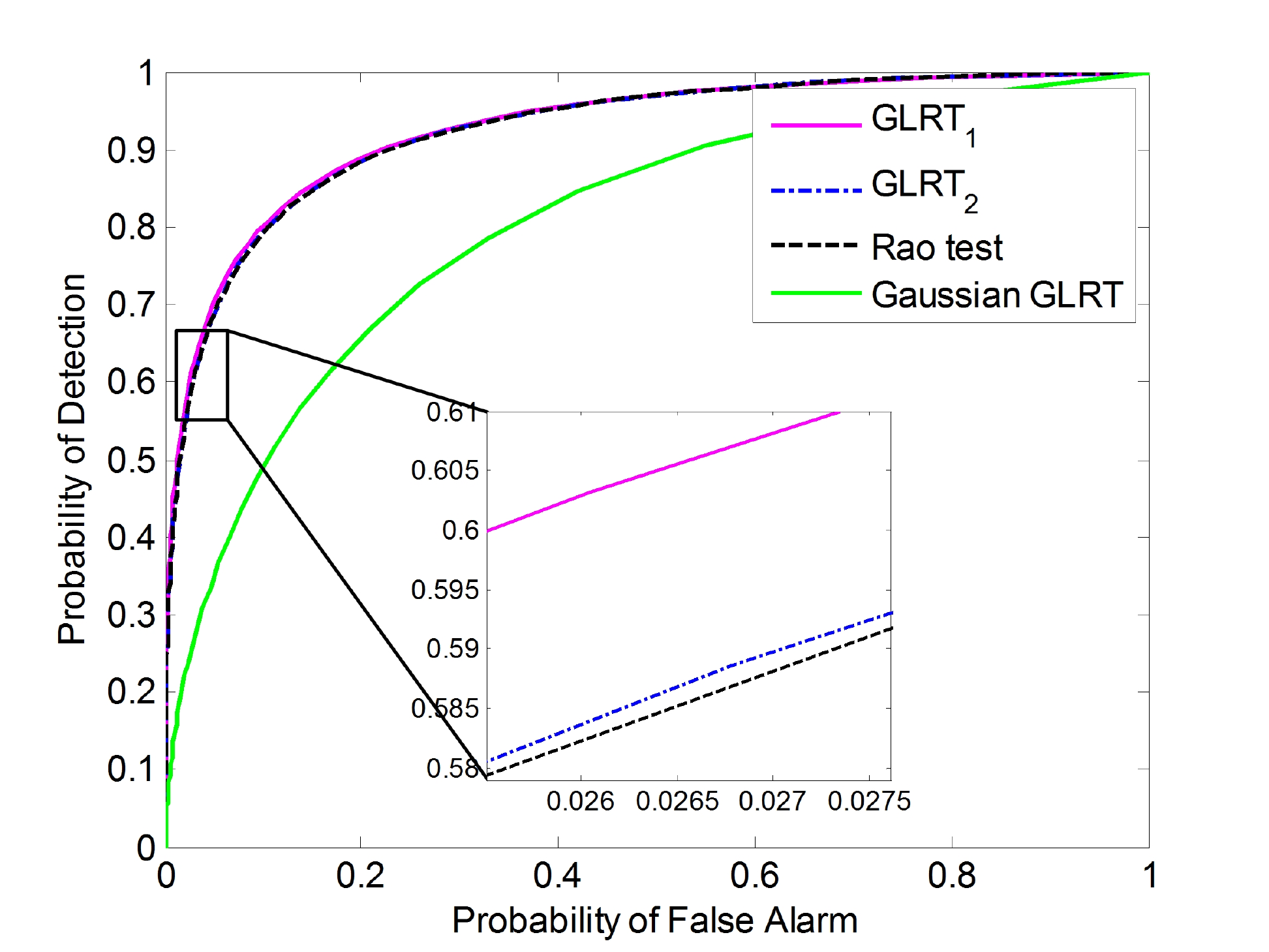}
\caption{Receiver operating characteristics of three test detectors with AR process for SAR signal detection: $\text{GLRT}_1$ is performed with unknown amplitude, $\text{GLRT}_2$ is used with unknown amplitude and Gaussian noise variance, and $\text{Rao}$ test is employed with unknown amplitude, Gaussian noise variance and AR coefficients, compared with the Gaussian GLRT detector.} 
\label{fm_roc}
\end{figure}

We first examined the case when $\sigma^2 = 1$. One example of the FM signal, noise and their observations are shown in Fig. \ref{fm_signal_noise}. In Fig. \ref{fm_signal_noise}(a), one can see the frequency of the FM signal is linearly modulated. In Fig. \ref{fm_signal_noise}(b), one example of white Gaussian noise and its colored noise filtered by an AR($p$) process is shown. A significant description difference between the white Gaussian noise and its AR($p$) colored noise can be seen. Such difference is further illustrated in Fig. \ref{fm_signal_noise}(c) where it is hard to identify whether a FM signal is presented in such observation. In this FM signal detection task, we apply $\text{GLRT}_1$ for the case of unknown amplitude, $\text{GLRT}_2$ for the case of unknown amplitude and unknown Gaussian variance, $\text{Rao}$ test for the case of unknown amplitude, variance and AR coefficients. Moreover, we compared with the Gaussian GLRT detector in \cite{Kay:1998_detection} where the unknown noise is modeled by a Gaussian distribution model. The Monte Carlo simulation results of these four detectors are shown in Fig. \ref{fm_roc}. The performance comparison in Fig. \ref{fm_roc} illustrates the agreement of our previous optimality analysis of these three test detectors regarding to the information loss. Meanwhile, it shows that the detection performance would be degraded if the colored Gaussian noise is incorrectly considered as the Gaussian noise. 

\begin{figure}[!ht]
\centering
\includegraphics[width=14cm]{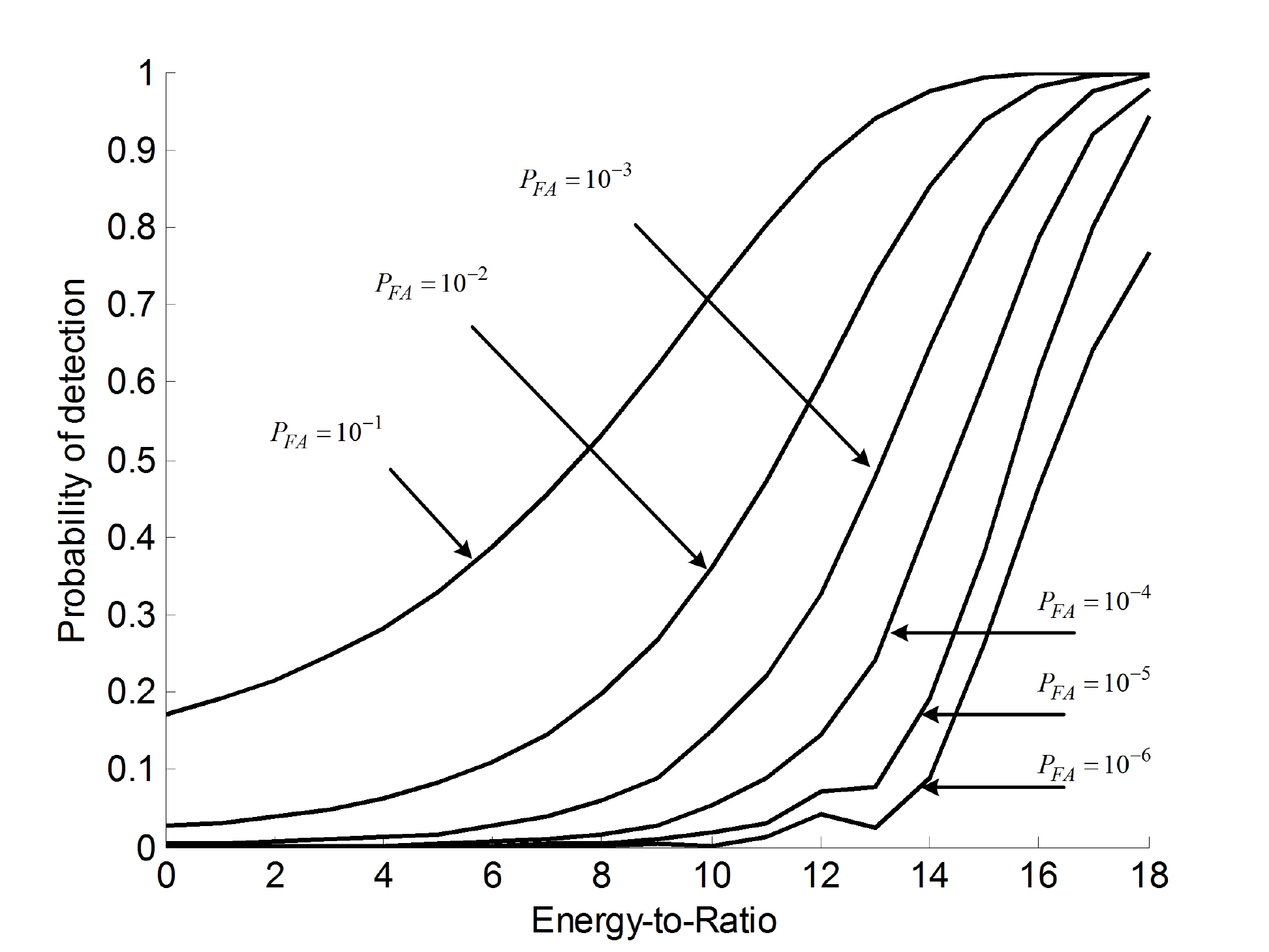}
\caption{GLRT detection performance for FM signal with unknown amplitude} 
\label{fm_signal_GLRT1}
\end{figure}

\begin{figure}[!ht]
\centering
\includegraphics[width=14cm]{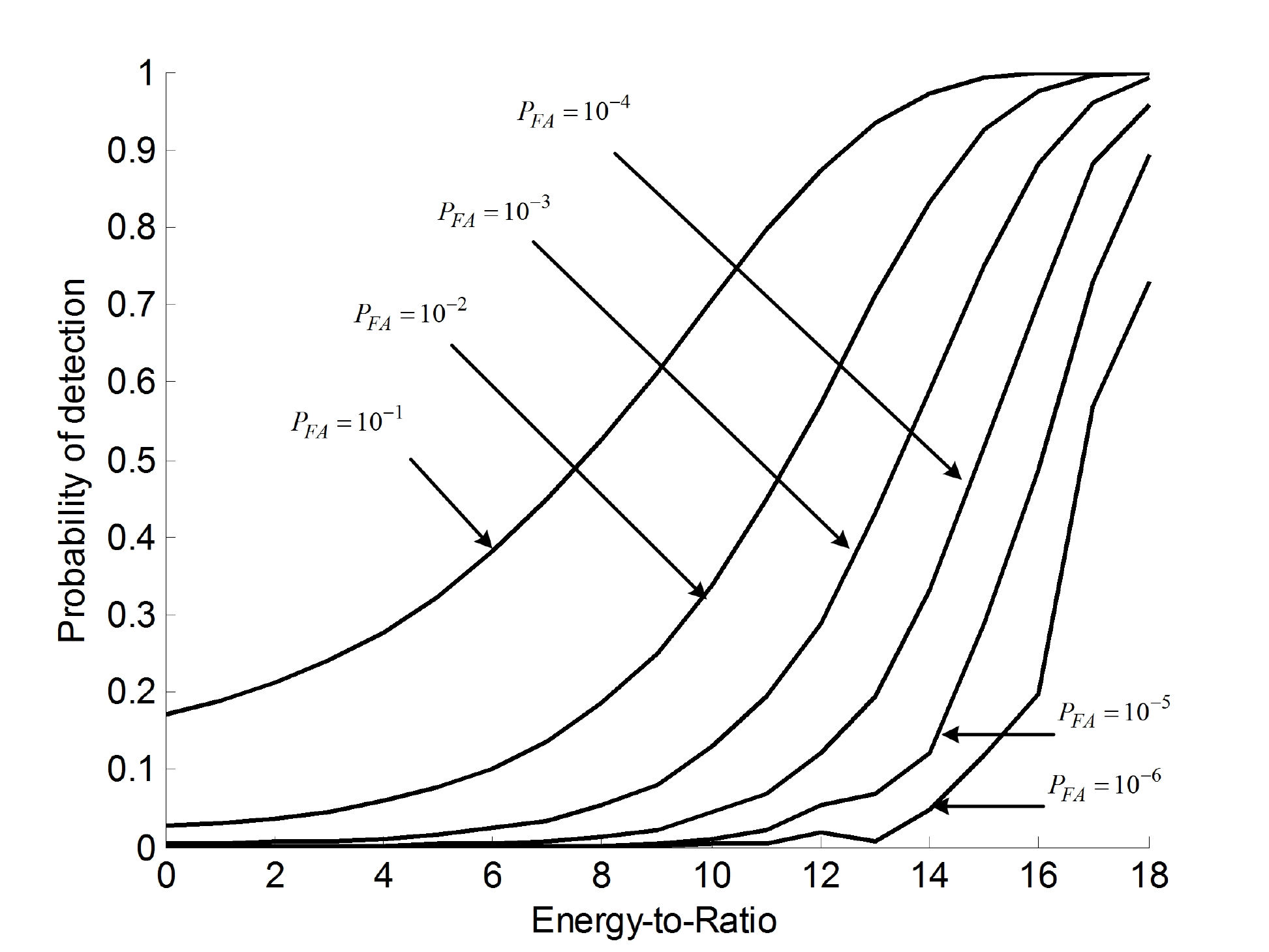}
\caption{GLRT detection performance for FM signal with unknown amplitude and Gaussian noise variance} 
\label{fm_signal_GLRT2}
\end{figure}

\begin{figure}[!ht]
\centering
\includegraphics[width=14cm]{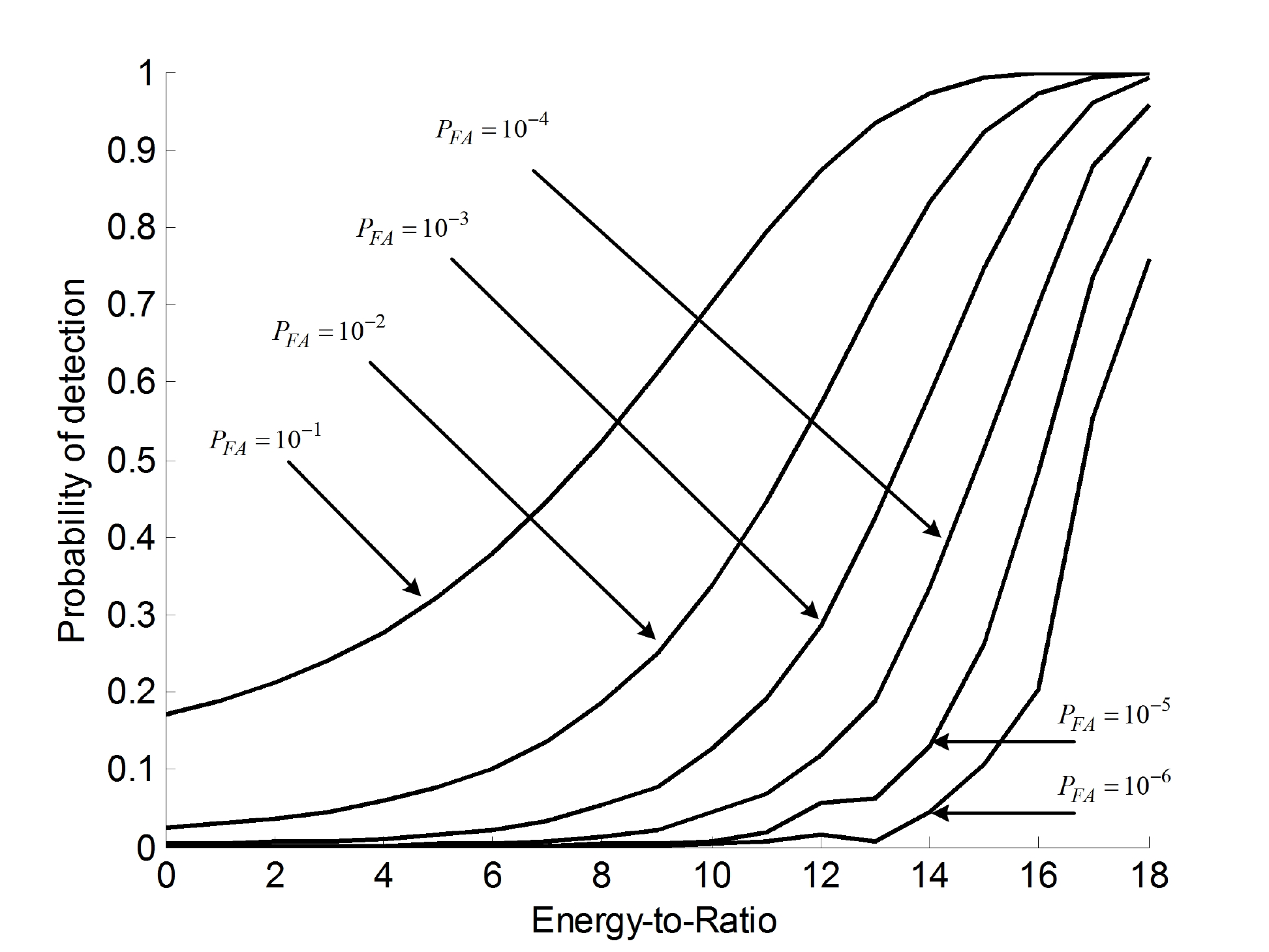}
\caption{Rao test detection performance for FM signal with unknown amplitude, Gaussian noise variance, and AR coefficients} 
\label{fm_signal_Rao}
\end{figure}

We next evaluated the influence of noise on detection performance of three detectors with different energy-to-noise ratio which is defined as the ratio of FM signal power to the AR($p$) colored noise power in dB. The probability of detection of three individual test detectors versus energy-to-noise ratio is given in Fig. \ref{fm_signal_GLRT1}, Fig. \ref{fm_signal_GLRT2} and Fig. \ref{fm_signal_Rao}, respectively. Given a specific probability of false alarm, the probability of detection for all the three detectors increases with the energy-to-noise ratio. Also, a high probability of false alarm usually leads to a high probability of detection, as the threshold of each test detector decreases. 

\subsection{Applications of Online Time Series Signal Detection}
We next apply three detectors for online time series object detection in the presence of colored noise. Assuming an object is moving across the area monitored by the radar or other sensors, we aim to detect the moving object from the observations that are disturbed with colored noise. For each time step, we have $n_r \times n_c$ observations for the monitored area, where $n_r$ denotes the number of rows and $n_c$ denotes the number of columns. At time $t_0$, we test whether an object appears at a specific location based on the observations from its previous $N$ time steps. This online moving object detection problem is formulated as the following binary hypothesis:
\begin{align}
&\mathcal{H}_0 : x_{ij}[n] = w_{ij}[n] & n = 0,1,\cdots,N-1 \nonumber \\
& \mathcal{H}_1 : x_{ij}[n] = s_{ij}[n-n_0]+ w_{ij}[n] & n = 0,1,\cdots,N-1
\end{align}
where $x_{ij}[n]$ denotes the signal observation at row $i$ and column $j$ at time $(t_0 + n - N + 1)$, $w_{ij}$ represents colored noise which is modeled by an AR($p$) process, and $s_{ij}$ denotes the moving object signal with the appearing time $n_0$ and the duration $l$ to be detected, which is given by
\begin{align}
\mathbf{s}_{ij} = A_{ij} [\underbrace{0, \cdots, 0}_{n_0}, \underbrace{1 \cdots 1}_{l}, 0, \cdots, 0]^T
\end{align}
This time series signal detection problem can be solved by using our detectors. However, unlike the hypothesis testing problem defined in Eq. \ref{classic_hypo}, two additional unknown parameters of moving object need to be examined for online moving object detection: the appearing time $n_0$ that the moving object appears in these $N$ time observations and its duration $l$. 

For these two additional unknown parameters, a GLRT decides $\mathcal{H}_1$ if 
\begin{align}
L_G(\mathbf{x}) = \frac{p(\mathbf{x};\hat{n}_0, \hat{l}, \mathcal{H}_1)}{p(\mathbf{x};\mathcal{H}_0)} > \gamma
\end{align}
where $\hat{n}_0$ and $\hat{l}$ are the MLEs of $n_0$ and $l$, respectively. Note that the PDF under $\mathcal{H}_0$ does not depend on $n_0$ and $l$, and the logarithm is a monotonic function. Thus, we can build a new GLRT detector when considering the two unknown parameters $n_0$ and $l$:
\begin{align}
T'_{G}(\mathbf{x}) & = \ln L_G(\mathbf{x}) = \ln \frac{p(\mathbf{x};\hat{n}_0, \hat{l}, \hat{\bfth}_1, \mathcal{H}_1)}{p(\mathbf{x}; \hat{\bfth}_0, \mathcal{H}_0)} \nonumber \\
& = \ln \frac{ \max_{l, n_0} p(\mathbf{x};n_0, l,  \hat{\bfth}_1, \mathcal{H}_1)}{p(\mathbf{x}; \hat{\bfth}_0, \mathcal{H}_0)} \nonumber \\
& =  \max_{l, n_0} \ln \frac{ p(\mathbf{x};n_0, l, \hat{\bfth}_1, \mathcal{H}_1)}{p(\mathbf{x}; \hat{\bfth}_0, \mathcal{H}_0)} \nonumber \\
& =  \max_{l, n_0} T_{G} (\mathbf{x}; n_0, l)
\end{align}
where $l \in [1, N]$ and $n_0 \in [0, N-l]$.

For the Rao test, in a similar fashion, we have
\begin{align}
T'_{R}(\mathbf{x}) = \max_{l,n_0} T_{R} (\mathbf{x};n_0,l)
\end{align}

Hence, the above three detectors with the unknown object appearing time $n_0$ and the unknown duration $l$ can be rewritten as
\begin{align}
\label{detectors_video}
& T'_{G_1}(\mathbf{x}) = \max_{l \in [1, N], n_0 \in [0, N-l]} T_{G_1}(\mathbf{x};n_0,l) \nonumber \\
& T'_{G_2}(\mathbf{x}) = \max_{l \in [1, N], n_0 \in [0, N-l]} T_{G_2}(\mathbf{x};n_0,l) \nonumber \\
& T'_{R}(\mathbf{x}) = \max_{l \in [1, N], n_0 \in [0, N-l]} T_{R}(\mathbf{x};n_0,l)
\end{align}

Given the current and previous $N$ time observations, denoted by $\mathbf{I}(t)_{t=0,1,\cdots,N-1}$, $\mathbf{I}(t) \in \mathcal{R}^{n}$ and $n = n_r n_c$, we examine each spatial position and test whether an object signal appears at current time. For the $(i,j)$ position, we denote the $N$ observations by $\mathbf{x} = [x_{ij}[0], x_{ij}[1],\cdots, x_{ij}[N-1]]^{T}$. Applying the above detectors in Eq. (\ref{detectors_video}), we obtain three target values $T'_{G_1}(\mathbf{x})$, $T'_{G_2}(\mathbf{x})$, $T'_{R}(\mathbf{x})$ and their corresponding parameters estimate of $\hat{n}_0$ and $\hat{l}$. For online moving object detection, we determine $\mathcal{H}_1$ if $\hat{n}_0 + \hat{l} = N$. 

\begin{figure}[!ht]
\centering
\includegraphics[width=14cm]{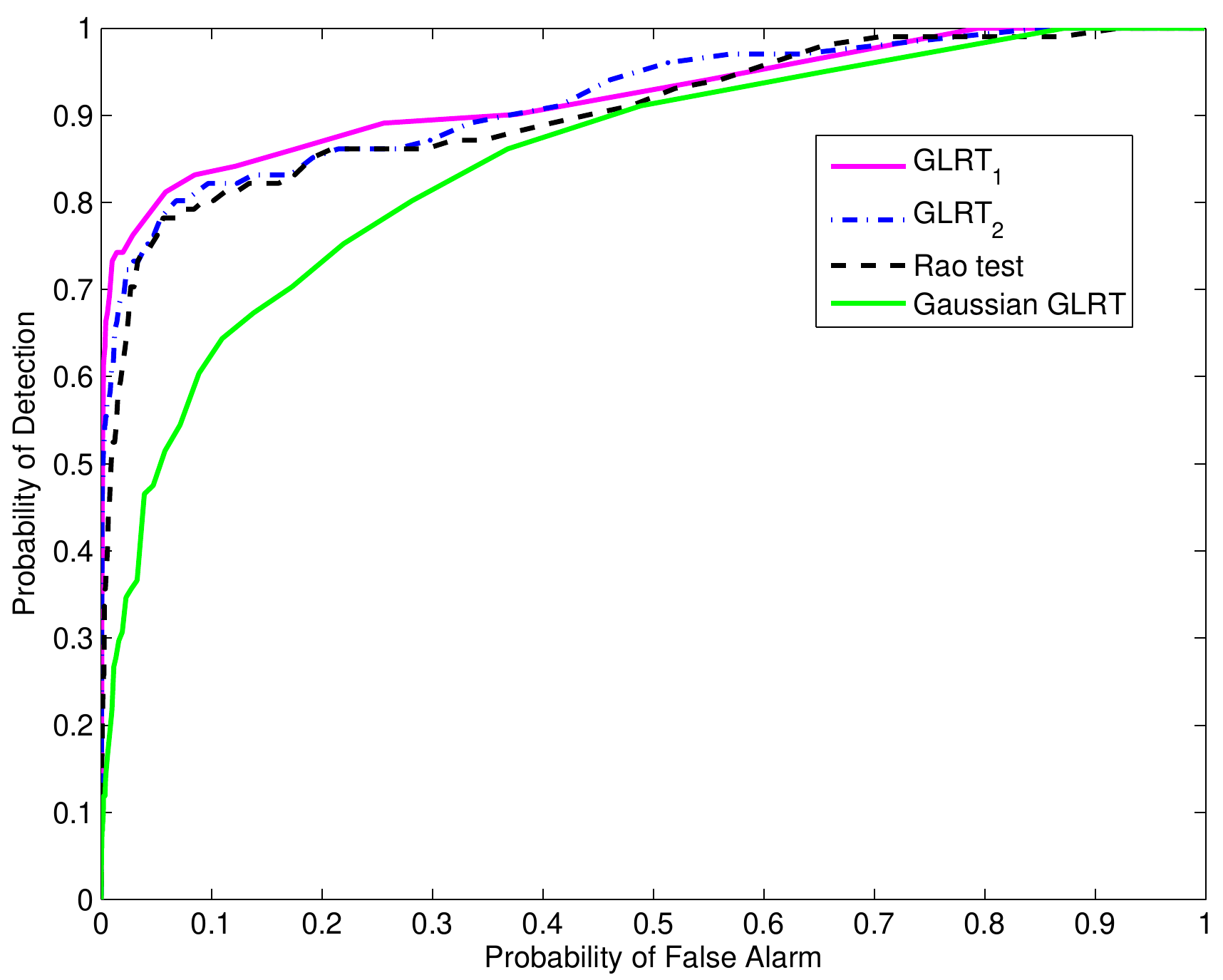}
\caption{Receiver operating characteristics of three test detectors with AR process for online moving object detection, compared with the Gaussian GLRT detector.} 
\label{ARSIMVideo}
\end{figure}

To evaluate the performance of detection, we simulated an object with amplitude $A = 1$ moving across the observed scene in the presence of colored noise. The colored noise is modeled by an AR($p$) process with coefficient factors $\alpha_1 = -0.2$ and $\alpha_2 = 0.8$ and the noise variance $\sigma^2 = 0.2$. The speed of moving object is $0.2 m/s$ and the duration that the moving object stays at one pixel is about $5$ seconds. For each new signal to detect, we use its $N = 20$ previous time as observations. As we aim to detect whether the object signal appears or not, we simplify the hypothesis as $\mathcal{H}_1: A \neq 0$ versus $\mathcal{H}_0: A = 0$. Thus, for the three detectors, we have $\mathbf{A} = \mathbf{I}$ where $\mathbf{I}$ is identity matrix, $\mathbf{b} = \mathbf{0}$, and 
\begin{align}
\mathbf{H} = [\underbrace{0, \cdots, 0}_{n_0}, \underbrace{1 \cdots 1}_{l}, \underbrace{0, \cdots, 0}_{N-n_0-l}]^T
\end{align}
The detection performance of our three test detectors in comparison with the Gaussian GLRT detector is shown in Fig. \ref{ARSIMVideo}. The detection results in Fig. \ref{ARSIMVideo} illustrate the optimality of three detectors in terms of information loss. On the other hand, the poorer detection performance of Gaussian GLRT detector indicates that the detection performance would be degraded, if the colored noise is modeled by a Gaussian distribution model. 

\subsection{Discussion}
As we discussed previously, the detection performance would decrease with the loss of information of distribution model. That is to say, the $T_{G_1}$ performs better than $T_{G_2}$, while $T_{G_2}$ performs better than $T_{R}$, if their corresponding parameters of distribution model are known. Although the coefficient factor of AR model $\bfalph$ and the variance of Gaussian noise $\sigma^2$ are difficult to know prior to the detection, they could be estimated in Eq. (\ref{eq:yw}) and (\ref{T2_estimatSig}) with some available sequences under $\mathcal{H}_0$. Because $\bfalph$ and $\sigma^2$ are nuisance parameters which are the same under both $\mathcal{H}_0$ and $\mathcal{H}_1$, their estimates under $\mathcal{H}_0$ can be applied to the detectors directly. Moreover, our simulation results demonstrate that the detection performance would decrease if the colored noise is incorrectly modeled by the Gaussian distribution model, which further illustrates the advantage of three presented detectors. 

\section{Conclusion}
This paper considers a general signal detection problem in the presence of colored noise. We generalize the signal detection problem by modeling the colored noise with an autoregressive process and present closed forms of three detectors using the GLRT and Rao test criterias when different parameters are unknown. The expressions of the unknown parameter estimates are given using the maximum likelihood methods, and the asymptotic distributions of these test detectors are also given. The detection performance and optimality of three detectors are studied through simulations. Two examples of signal detection imply a wide application of test detectors. 

\section{Acknowledgment}
This research was partially supported by National Science Foundation (NSF) under grant ECCS 1053717 and CCF 1439011, and the Army Research Office under grant W911NF-12-1-0378.

\section{APPENDIX 1: Proof of Theorem 1}
\label{Appendix 1}
Assume the data satisfies the classic AR model, that is
\begin{align}
\mathbf{x} = \mathbf{H} \bfth + \mathbf{w}
\end{align}
and the noise $\mathbf{w}$ satisfies the AR model,
\begin{align}
w[n] = \sum_{i=1}^p \alpha_i w[n-i] + v[n] \quad \quad n = 0, 1, \cdots, N-1
\end{align}
Given $x[-1], x[-2], \cdots, x[-p]$, we have
\begin{align}
& p(\mathbf{x}, \bfth) = \prod_{n=0}^{N-1} p(x[n] | x[n-1], \cdots, x[n-p]) \nonumber \\
& = \frac{1}{(2 \pi \sigma^2)^{N/2}} \exp \left\{ -\frac{1}{2\sigma^2} \sum_{n=0}^{N-1} \left[ s[n] - \sum_{k=1}^p \alpha_k s[n-k] \right] \right\}
\end{align}
where $s[n]$ denotes the $n$-th element of vector $\mathbf{s} = \mathbf{x} - \mathbf{H} \bfth$. The term in exponential function can be reduced as 
\begin{align}
\sum_{n=0}^{N-1} \left[ s[n] - \sum_{k-1}^p \alpha_k s[n-k] \right] = (\mathbf{T} \mathbf{s} + \mathbf{c})^T(\mathbf{T} \mathbf{s} + \mathbf{c})
\end{align}
when $\mathbf{T}$ and $\mathbf{c}$ are written as Eq. (\ref{T_c_form}).

Moreover, we have
\begin{align}
& (\mathbf{T} \mathbf{s} + \mathbf{c})^T(\mathbf{T} \mathbf{s} + \mathbf{c}) = (\mathbf{T} (\mathbf{x} - \mathbf{H} \bfth) + \mathbf{c})^T(\mathbf{T} (\mathbf{x} - \mathbf{H} \bfth) + \mathbf{c}) \nonumber \\
& = (\mathbf{T} \mathbf{x} + \mathbf{c} - \mathbf{T} \mathbf{H} \bfth)^T(\mathbf{T} \mathbf{x} + \mathbf{c} - \mathbf{T} \mathbf{H} \bfth)
\end{align}
By setting $\mathbf{x}' = \mathbf{T} \mathbf{x} + \mathbf{c}$ and $\mathbf{H}' = \mathbf{T} \mathbf{H}$, we have
\begin{align}
\label{Approximate 1_x'distribution}
& p(\mathbf{x}, \bfth) = p(\mathbf{x}', \bfth) \nonumber \\
& = \frac{1}{(2 \pi \sigma^2)^{N/2}} \exp \left\{ -\frac{1}{2\sigma^2} (\mathbf{x}' - \mathbf{H}' \bfth)^T(\mathbf{x}' - \mathbf{H}' \bfth) \right\}
\end{align}
which means the new variable $\mathbf{x}'$ satisfies $\mathbf{x}' \sim \mathcal{N}(\mathbf{H}' \bfth, \sigma^2 \mathbf{I})$, where only the signal $\bfth$ is unknown. Through the linear transformation, the classic AR model can be transformed as the linear Gaussian model for which Kay in \cite{Kay:1998_detection} gave the GLRT detector. Similar in \cite{Kay:1998_detection}, the test statistic of GLRT has the form of
\begin{align}
\label{T1_xhat}
& T_{GLRT}(\mathbf{x}') = 2\ln L_G(\mathbf{x}') \nonumber \\
& =  \frac{( \mathbf{A} \hat{\bfth}_1 - \mathbf{b})^{T} [ \mathbf{A} [ (\mathbf{H}')^{T} \mathbf{H}' ]^{-1} \mathbf{A}^{T} ]^{-1} ( \mathbf{A}\hat{\bfth}_1 - \mathbf{b})}{\sigma^2}  > \gamma'
\end{align}
where $\hat{\bfth}_1$ is the MLE of $\bfth$ under $\mathcal{H}_1$. $\hat{\bfth}_1$ is equivalent to the unconstrained MLE of $\hat{\bfth}$, which is 
\begin{align}
\label{T1_theta_xhat}
\hat{\bfth}_1 = (\mathbf{H}'^T \mathbf{H}')^{-1} \mathbf{H}'^T \mathbf{x}'
\end{align}

Because the GLRT detector decides $\mathcal{H}_1$ if 
\begin{align}
L_G(\mathbf{x}) = \frac{p(\mathbf{x};\hat{\bfth}_1)}{p(\mathbf{x};\hat{\bfth}_0)} > \gamma
\end{align} 
and $p(\mathbf{x}, \bfth) = p(\mathbf{x}', \bfth)$, we have
\begin{align}
L_G(\mathbf{x}) = \frac{p(\mathbf{x}';\hat{\bfth}_1)}{p(\mathbf{x}';\hat{\bfth}_0)} > \gamma
\end{align}
where $\hat{\bfth}_1$ and $\hat{\bfth}_0$ are the MLEs of $\bfth$ under $\mathcal{H}_1$ and $\mathcal{H}_0$, respectively, so that we replace $\mathbf{x}' = \mathbf{T} \mathbf{x} + \mathbf{c}$ and $\mathbf{H}' = \mathbf{T} \mathbf{H}$ in Eq. (\ref{T1_xhat}) and Eq. (\ref{T1_theta_xhat}) and obtain the the GLRT detector for $\mathbf{x}$ with the form of Eq. (\ref{T1}). 

Also, the exact performance can be derived easily with the same way in \cite{Kay:1998_detection}. Since $\mathbf{x}' \sim \mathcal{N}(\mathbf{H}'\bfth, \sigma^2 \mathbf{I})$ and $\hat{\bfth}_1$ is the linear transformation of $\mathbf{x}'$ in Eq. (\ref{T1_theta_xhat}), we have
\begin{align}
\hat{\bfth}_1 \sim \mathcal{N}(\bfth, \sigma^2(\mathbf{H}'^T \mathbf{H}')^{-1})
\end{align}
and 
\begin{align}
\mathbf{A}\hat{\bfth}_1 - \mathbf{b} \sim \mathcal{N}(\mathbf{A}\bfth - \mathbf{b}, \sigma^2 \mathbf{A}(\mathbf{H}'^T \mathbf{H}')^{-1}\mathbf{A}^T)
\end{align}
Since the test statistic of GLRT has form of $(\mathbf{A}\hat{\bfth}_1 - \mathbf{b})^T C^{-1} (\mathbf{A}\hat{\bfth}_1 - \mathbf{b})$, we have
\begin{align}
T_{GLRT}(\mathbf{x}) \sim \left\{ \begin{array}{ll}
\chi_r^2  & \text{ under } \mathcal{H}_0 \\
\chi_r'^2 (\lambda) & \text{ under } \mathcal{H}_1
\end{array}
\right.
\end{align}
where $\lambda$ has the form in Eq. (\ref{T1_lamda}). Thus, we have the exact performance shown in Eq. (\ref{Theory 1 Performance}).

\section{APPENDIX 2: Proof of Theorem 3}
\label{Appendix 2}
From Eq. (\ref{Approximate 1_x'distribution}), we know that $\mathbf{x}' \sim \mathcal{N}(\mathbf{H}'\bfth_r, \sigma^2 \mathbf{I})$. According to the the theorem of Cramer-Rao Lower Bound in \cite{Kay:1993_estimation}, the consistent unrestricted MLE of $\hat{\bfth}_r$ satisfies 
\begin{align}
\frac{\partial \ln p \left( \mathbf{x}'; \bfth \right)}{\partial \bfth_r} = \mathbf{I}( \bfth_r ) ( \hat{\bfth}_r - \bfth_r )
\end{align}
Similar in Eq. (\ref{T1_theta_xhat}), we can obtain the exact form of $\hat{\bfth}_r$
\begin{align}
\hat{\bfth}_r = ( \hat{\mathbf{H}}'^T \hat{\mathbf{H}}' )^{-1} \hat{\mathbf{H}}'^T \mathbf{x}'
\end{align}
where $\hat{\mathbf{H}}' = \hat{\mathbf{T}} \mathbf{H}$ and $\mathbf{x}' = \hat{\mathbf{T}}\mathbf{x} + \hat{\mathbf{c}}$. $\hat{\mathbf{T}}$ and $\hat{\mathbf{c}}$ are the MLEs of $\mathbf{T}$ and $\mathbf{c}$, respectively, with the form of Eq. (\ref{Theory3: T_c_hat_form}) obtained by solving the Yule-Walker equation in Eq. (\ref{eq:yw}). Furthermore, we have
\begin{align}
\left. \frac{\partial \ln p(\mathbf{x}; \bfth)}{\partial \bfth_{r}} \right |^{T}_{\bfth = \hat{\bfth}_0} = \frac{1}{\hat{\sigma}^2} \hat{\mathbf{H}}'^T \hat{\mathbf{H}}' (\hat{\bfth}_r - \bfth_{r_0})
\end{align}
\begin{align}
\mathbf{I}(\bfth_r) = \frac{\hat{\mathbf{H}}'^T\hat{\mathbf{H}}'}{\hat{\sigma}^2}
\end{align}
where $\hat{\sigma}^2$ is the MLE of $\sigma^2$ under $\mathcal{H}_0$ with the form of Eq. (\ref{eq:Theory3_sigma_hat}). Thus, we have
\begin{align}
T_R(\mathbf{x}') & = \frac{1}{\hat{\sigma}^2} (\hat{\bfth}_r - \bfth_{r_0})^T \hat{\mathbf{H}}'^T \hat{\mathbf{H}}' (\hat{\mathbf{H}}'^T \hat{\mathbf{H}}')^{-1} \hat{\mathbf{H}}'^T \hat{\mathbf{H}}'(\hat{\bfth}_r - \bfth_{r_0}) \nonumber \\
& = \frac{1}{\hat{\sigma}^2} (\hat{\bfth}_r - \bfth_{r_0})^T \hat{\mathbf{H}}'^T \hat{\mathbf{H}}' (\hat{\bfth}_r - \bfth_{r_0}) \nonumber \\
& = \frac{1}{\hat{\sigma}^2} [\mathbf{x}'^T \hat{\mathbf{H}}' (\hat{\mathbf{H}}'^T \hat{\mathbf{H}}')^{-1} \hat{\mathbf{H}}'^T \mathbf{x}' - \mathbf{x}'^T\hat{\mathbf{H}}'\bfth_{r_0} \nonumber \\
& \quad - \bfth_{r_0}^T\hat{\mathbf{H}}'^T\mathbf{x}' + \bfth_{r_0}^T\hat{\mathbf{H}}'^T \hat{\mathbf{H}}'\bfth_{r_0}]
\end{align}
Hence, because of $\mathbf{x}' = \mathbf{T} \mathbf{x} + \mathbf{c}$, if we assume $\mathbf{A}$ is invertible and $\bfth_{r_0} = \mathbf{A}^{-1} \mathbf{b}$, the Rao test for the unknown signal and unknown nuisance parameters is written as
\begin{align}
T_R(\mathbf{x}) & = \frac{1}{\hat{\sigma}^2} [(\hat{\mathbf{T}} \mathbf{x} + \mathbf{c})^T \hat{\mathbf{T}} \mathbf{H} ((\hat{\mathbf{T}} \mathbf{H})^T \hat{\mathbf{T}} \mathbf{H})^{-1} (\hat{\mathbf{T}} \mathbf{H})^T (\hat{\mathbf{T}} \mathbf{x} + \mathbf{c}) \nonumber \\
& - (\hat{\mathbf{T}} \mathbf{x} + \mathbf{c})^T \hat{\mathbf{T}} \mathbf{H} \mathbf{A}^{-1} \mathbf{b} - (\mathbf{A}^{-1} \mathbf{b})^T (\hat{\mathbf{T}} \mathbf{H})^T (\hat{\mathbf{T}} \mathbf{x} + \mathbf{c}) \nonumber \\
& + (\mathbf{A}^{-1} \mathbf{b})^T (\hat{\mathbf{T}} \mathbf{H})^T \hat{\mathbf{T}} \mathbf{H} \mathbf{A}^{-1} \mathbf{b}]
\end{align}
It is easily demonstrate that the asymptotic performance of Rao test given in Eq. (\ref{eq:Theory3_performance}) is the same as the GLRT, because Rao test is equivalent to the GLRT under the assumptions that the data size is large, that is $N \rightarrow \infty$, and the signal $\bfth_{r}$ is either $\bfth_{r_0} = \mathbf{A}^{-1}\mathbf{b}$ under $\mathcal{H}_0$ or near $\bfth_{r_0}$ under $\mathcal{H}_1$ \cite{Kay:1998_detection}. 


\bigskip

\section*{Reference}

\bibliographystyle{elsarticle-num} 
\bibliography{eefref}

\end{document}